\documentclass[letterpaper,10pt]{article}
\usepackage{multicol}
\usepackage{graphicx}
\begin{document}
\begin{center}
{\bf\large Classical and quantum chaos
	in a circular billiard with a straight cut}\\
\vskip 3ex
	{Suhan Ree and L.E. Reichl}\\
	{\em Center for Studies in Statistical Mechanics and Complex Systems}\\ 
	{\em The University of Texas at Austin, Austin, Texas 78712}\\
	(July 6, 1998) 
\vskip 2ex

\parbox{14cm}{\indent\indent
We study classical and quantum dynamics of a particle in a circular billiard
with a straight cut.
This system can be integrable, nonintegrable with {\em soft} chaos,
or nonintegrable with {\em hard} chaos, as we vary the size of the cut.
We use a quantum web to show differences in the quantum manifestations of
classical chaos for these three different regimes.
}
\end{center}
\vskip 5ex
\begin{multicols}{2}
%------Sec. I---------------------------------------
\section*{I. Introduction}
\indent\indent
In recent years, the dynamics of noninteracting particles in
two-dimensional (2D)
billiards has been studied in many different billiard shapes [1-6].
In this paper, we study the classical and quantum dynamics of a particle in a
 closed circular billiard with a straight cut
(we name this system the ``Moon" billiard; see Fig.~1).
The classical system can exhibit both integrable
and nonintegrable behavior.
It also shows two distinct types of chaotic behavior, both ``hard chaos" and
``soft chaos"
as we change the size of the cut.
Systems whose dynamics consists of a mixture of stable and
unstable periodic orbits are said to exhibit {\em soft} chaos. If all
periodic orbits in a system
are unstable throughout,  the behavior is
called {\em hard} chaos [7].

The quantum version of these billiards has a discrete energy
spectrum,
and chaos (or nonintegrability) manifests itself in the quantum billiard in
indirect ways.
The statistics of energy levels
for classically integrable systems is different from that for classically
chaotic systems [2-5,8-11], and one commonly uses
the spectral statistics of energy levels as
signature of an underlying chaos.
Also discrete symmetries of the system should be handled carefully.

In this paper, we focus on a diagnostic tool which has proven successful
for spin systems.  We calculate the quantum web [11-14]
using about a hundred lowest-energy eigenstates,
and observe patterns for classically different cases.
The lattice-structure, which appears for classically integrable cases,
breaks in different ways for cases with soft chaos and hard chaos.
The Husimi function [15,16], which extracts the quantum Poincar\'{e} section
from a quantum state, is also calculated for some selected
energy eigenstates to examine the quantum web more closely.

%----------------------------------------------------------
\section*{II. The Classical Billiard}
\indent\indent
In this section, we focus on the classical dynamics of the Moon
billiard.
We introduce a dimensionless parameter $w=W/R$
to characterize the system 
where $W$ is the width of the billiard and $R$ is the radius
(see Fig.~1).
Then we define
$M_w$ as a circular disk with a straight cut with $w$.
For example, $M_2$ is a full circle and
$M_1$ is a half circle, and in general $w$ is in the
range of $0<w\le2$.
The classical Hamiltonian of a particle inside $M_w$,
using polar coordinates $(r,\theta)$, is
\begin{equation}
H_w(r,\theta,p_r,p_\theta)={p_r^2\over 2m}+{p_\theta^2\over 2mr^2}
+{V^R_w(r,\theta)},
%\eqno(1)
\end{equation}
where $V^R_w(r,\theta)$ represents the hard-wall potential of
the billiard with radius $R$ and $W=wR$.
To begin, we will study
the full circle ($M_2$) and the half circle
($M_1$). Both of cases are integrable.
Then nonintegrable cases will be examined.

\subsection*{A. Full-Circle Billiard, $M_2$}
\indent\indent
The dynamics of a particle inside $M_2$ is integrable because
there are two constants of motion, the energy $E$ and the angular momentum
$p_\theta$. Given $E$ and $p_\theta$, the orbit lies on a 2D torus
(2-torus) in phase space.
There exists a canonical
transformation to action-angle variables, $(J_r,J_\theta,\phi_r,
\phi_\theta)$, where actions are given by
\begin{equation}
	\begin{array}{rcl}
	J_r&=&{\sqrt{2mE}\over\pi}\left[\sqrt{R^2-{p_\theta^2
                      \over2mE}}-{|p_\theta|\over\sqrt{2mE}}\,\cos^{-1}
                        \left({|p_\theta|\over R\sqrt{2mE}}\right)\right],\\
        J_\theta&=&p_\theta,
	\end{array}
%\eqno(2)
\end{equation}
with the new Hamiltonian $H'=E(J_r,J_\theta)$.
We can also find the angle variables, but it should be noted that $\phi_\theta$
is not equal to $\theta$.
We can find explicit expressions
of angular frequencies ${\dot \phi}_r=\omega_r$ and
${\dot \phi}_{\theta}=\omega_\theta$ as functions
of $E$ and $p_\theta$ using Eq.~(2),
\begin{equation}
\omega_r=\dot\phi_r={\partial E\over\partial J_r}
        =\left({\partial J_r\over\partial E}\right)^{-1}
        ={2\pi E\over\sqrt{2mER^2-p_\theta^2}},
%\eqno(3)
\end{equation}
\begin{eqnarray}
\omega_\theta&=&\dot\phi_\theta={\partial E\over\partial J_\theta}
        =-\left({\partial J_r\over\partial J_\theta}\right)
           \left({\partial J_r\over\partial E}\right)^{-1} \nonumber\\
        &=&{{\rm sgn}(p_\theta)\over\pi}~\omega_r(E,p_\theta)~\cos^{-1}\left(
        {|p_\theta|\over R\sqrt{2mE}}\right).
%\eqno(4)
\end{eqnarray}

It is useful to look at periodic orbits (orbits for which the
ratios of two angular frequencies are rational). We classify periodic
orbits using the notation $(m,n)$ where $m$ and $n$ are relatively
prime positive integers defined by
\begin{equation}
\left|{\omega_\theta\over\omega_r}\right|={1\over\pi}~\cos^{-1}
\left({|p_\theta|\over R\sqrt{2mE}}\right)={m\over n},
%\eqno(5)
\end{equation}
where $2m<n$ [see Fig.~2(a)].
If we have a periodic orbit classified as $(m,n)$, an
infinite number of rotated periodic orbits also belongs to $(m,n)$.
Therefore  periodic orbits in the full circle are {\em non-isolated} [1].
For any periodic orbit classified as $(m,n)$, there are $n$ bounces and
$m$ revolutions in one cycle. On the other hand, a non-periodic orbit
will not come back to the starting point, and eventually fill the whole
2-torus.

\subsection*{B. Half-Circle Billiard, $M_1$}
\indent\indent
For the half circle, we still have two constants of motion, $E$ and
$p_\theta^2$.
The range of $\theta$ is reduced to a half, $-{\pi\over2}<\theta<{\pi\over2}$,
but we can still construct 2-tori on which orbits lie.
For any orbit in $M_1$ there is a corresponding orbit in $M_2$.
(If there is an orbit in $M_2$, folding $M_2$ in half gives us a
corresponding orbit in $M_1$.)
Hence we can use the results found for $M_2$ to describe some periodic orbits.
A periodic orbit
is classified as $(m,n)'$ when the corresponding orbit in $M_2$ is
$(m,n)$.
Unlike for $M_2$, periodic orbits in $M_1$ are {\em isolated}, since there is no
rotational
symmetry.
In the group $(m,n)'$, there are an infinite number of different periodic orbits,
but only a few play an important role when $w$ is slightly less than 1.
In Fig.~2(b), those periodic orbits are shown.
The stabilities of these periodic orbits
are all neutral (neither stable nor unstable) like those in $M_2$.

It is useful to see periodic orbits in the space of $E$ and $|p_\theta|$.
A point in this space corresponds to a group of orbits with constants of motion
$E$ and $p_\theta^2$. When Eq.~(5) is satisfied, the group consists of
periodic orbits. Thus the condition to have periodic orbits is given by
Eq.~(5), and is plotted in Fig.~3.
They are densely populated in the classically allowed region. The classically
forbidden region in this space is given by $E-p_\theta^2/(2mR^2)<0$.
 In integrable cases, orbits are a point (zero-dimensional) in
$(E,|p_\theta|)$-space since $E$ and
$p_\theta^2$ are conserved, but as
the rotational symmetry breaks by changing $w$, $p_\theta^2$ is no longer
conserved
and orbits become one-dimensional.
We will see later that
Eq.~(5) gives us the condition for primary resonances
when $w$ is slightly less than 1.

\subsection*{C. $M_w$ when $0<w<1$ or $1<w<2$}
\indent\indent
Except two cases studied above ($w=1,2$), the system is nonintegrable, because
$E$ is an only constant of motion in the system of two degrees of freedom.
The stabilities of periodic orbits in $M_w$ give us one way to understand the
dynamical behavior
of the system. The simplest periodic orbit for any $w$ is an orbit with two
bounces,
going back and forth [for example, $(1,2)'$ in Fig.~2(b)].
To calculate the stability of this orbit, we need to imagine a new billiard
$M_w^2$, which is a composite of circular parts of two $M_w$'s facing each
other.
It is easy to see that $M_1^2$ is just $M_2$.
Orbits in $M_w$ have the correspondence with orbits in $M_w^2$ as orbits in
$M_1$ correspond to orbits in $M_2=M_1^2$.
The stability of this two-bounce orbit in $M_w^2$ can be calculated from a
$2\times2$-matrix ${\bf m}$, acting on $(\delta\theta,{\delta p_\theta\over 
R\sqrt{2mE}})$
on the boundary,
which decides the deviation after two bounces [1]
\begin{equation}
{\bf m}=\left(\matrix{2(2w-1)^2-1 & 4w(1-2w) \cr
            4(2w-1)(1-w)& 2(2w-1)^2-1\cr}\right).
%\eqno(6)
\end{equation}
The eigenvalues of ${\bf m}$, $\lambda_\pm$, are given in terms of
 the trace of ${\bf m}$,
\begin{equation}
\lambda_\pm={1\over2}\left\{ {\rm Tr}\,{\bf m} \pm
        \left[({\rm Tr}\,{\bf m})^2-4\right]^{1/2}\right\},
%\eqno(7)
\end{equation}
where we used ${\rm det}\,|{\bf m}|=1$ since ${\bf m}$ is area-preserving.
The orbit is neutral when $|{\rm Tr}\,{\bf m}|=2$,
stable when $|{\rm Tr}\,{\bf m}|<2$, and unstable when $|{\rm Tr}\, {\bf m}|>2$.
The two-bounce orbit is neutral when $w=0.5$, $w=1$, or $w=2$,
stable when $0<w<0.5$ or $0.5<w<1$, and unstable when $1<w<2$.
We have seen that all periodic orbits
are neutral when the system is integrable ($w=1,2$) but $w=0.5$ is
a special case as we will see later. From this result, we can predict that
there are no stable periodic orbits in the billiard $M_{1<w<2}$, and
periodic orbits
in $M_{0<w<1}$ are either stable or unstable, except when $w=0.5$.
(Ergodicity of the billiard $M_{1<w<2}$
has been proven by Bunimovich [17].)
The system shows hard chaos when $1<w<2$, and soft chaos when
$0<w<1$.

The Poincar\'{e} surface of section (PSS) is one way to observe the chaos.
Here PSS is a  two-dimensional surface $(\theta,p_\theta)$
at $r=R$ along the circular boundary ($-\theta_{max}<\theta<\theta_{max}$,
$-R\sqrt{2mE}<p_\theta<R\sqrt{2mE}$).
Each orbit gives a point in $(\theta,p_{\theta})$-space every time it
touches this surface.
Therefore PSS becomes a 2D area-preserving map.
In fully chaotic (ergodic) cases, points generated by an orbit will fill
out the
whole allowed space in the PSS.
In cases of soft chaos, we observe some structure. Some orbits
generate island chains and some orbits will fill some remaining regions in
a chaotic manner.
Since the KAM theorem [7,11] does not apply in this system due to
singularities of the boundary,
the existence of KAM tori seperating island chains is not guaranteed even for
small $\delta$ when $w=1-\delta$.

In Fig.~4, we show PSS's for various $w$'s.
In Fig.~4(a), one orbit is filling the whole region when $w=1.01$. This is a
fully chaotic case.
Figure 4(b) is an integrable case when $w=1$.
When $w=0.99$ [Fig.~4(c)],
we see island chains centered at the positions of periodic orbits along
with stochastic diffusion in the remaining region.
The resonance
condition, Eq.~(5), gives us locations of island chains. For example,
$(1,2)'$ gives us the location $p_\theta/(R\sqrt{2mE})=0$, $(1,4)'$
gives us $p_\theta/(R\sqrt{2mE})=\pm1/\sqrt{2}$, $(1,3)'$ gives us $p_\theta/
(R\sqrt{2mE})=\pm0.5$, and $(1,6)'$ gives us $p_\theta/(R\sqrt{2mE})=\pm\sqrt{3}/2$,
and so on. As $w$ decreases [Figs.~4(d) and 4(e)], we see increasing chaotic region
and some remaining island chains. When $w=0.5$ [Fig.~4(f)], periodic orbits
that have neutral stabilities reside in two axes, ($\theta=0$)-axis and
($p_\theta=0$)-axis. One orbit starting near the center is filling out
almost all region.
This is the most chaotic case when $0<w<1$.
As $w$ decreases further, we see bigger regular regions in PSS
in Figs.~4(g) and 4(h).

%--------------------------------------------------------------
\section*{III. The Quantum Billiard}
\indent\indent
In this section, we study the quantized $M_w$-billiard using a
quantum-web analysis with about a hundred lowest-energy eigenstates,
and we also look at some individual energy eigenstates using Husimi plots.

\subsection*{A. Numerical Method}
\indent\indent
The Schr\"{o}dinger equation for this 2D $M_w$-billiard is the
Helmholtz equation,
\begin{equation}
(\nabla^2+k^2)\Psi({\bf r})=0,
%\eqno(8)
\end{equation}
with the Dirichlet boundary condition, $\Psi=0$, on the boundary of $M_w$,
$B_w\equiv\partial M_w$, where $k^2=2mE/\hbar^2$ and
$\nabla^2={\partial^2\over\partial r^2}
+{1\over r}{\partial\over \partial r}+{1\over
r^2}{\partial^2\over\partial\theta^2}$
using polar coordinates.

For the classically integrable cases, this equation can be solved analytically.
The Hamiltonian $\hat{H_w}$ ($w=1$ or 2) and
the angular momentum $\hat{ p_\theta}$ ($\hat{p_\theta}^2$ for the half
circle) commute.
They are  generators of continuous symmetry transformations, the time
translation
and the rotation.
For a full circle $M_2$, 
we can find energy eigenstates which are simultaneous eigenstates of
$\hat{H_2}$ and $\hat{p_\theta}$,
\begin{equation}
\langle {\bf r}|l,k\rangle^{(f)} \propto J_l\left({\alpha_{lk}r\over R}\right)\,
e^{il\theta},
%\eqno(9)
\end{equation}
where $k$ is a positive integer, $l$ is an integer, and
$\alpha_{lk}$ is the $k$th zero of the  the Bessel function  $J_l(x)$.
And energy levels are given by
$E_{lk}^{(f)}=\hbar^2\alpha_{lk}^2/(2mR^2)$.
There exist two-fold degeneracies when $l\ne0$ since the system also has the parity
symmetry and $[\hat{p_\theta},\hat P]\ne0$ ($\hat P$ is the parity operator).
Then we can find another set of energy eigenstates, simultaneous eigenstates of
$\hat{H_2}$, $\hat{p_\theta}^2$, and $\hat P$,
\begin{equation}
\begin{array}{rcl}
\langle {\bf r}|l,k,+\rangle^{(f)} &\propto& J_l\left({\alpha_{lk}r\over
R}\right)\,
        \cos(l\theta)~~~(l\ge0),\\
\langle {\bf r}|l,k,-\rangle^{(f)} &\propto& J_l\left({\alpha_{lk}r\over
R}\right)\,
        \sin(l\theta)~~~(l\ge1),
%\eqno(10)
\end{array}
\end{equation}
where $-\pi<\theta<\pi$.
The latter will be used in the calculation of the quantum web.

For a half circle $M_1$, energy levels are the same as those of the
full circle without levels with $l=0$, and the energy eigenfunctions are
\begin{eqnarray}
\begin{array}{rcl}
\langle {\bf r}|l,k,+\rangle^{(h)} &\propto& J_l\left({\alpha_{lk}r\over
R}\right)\,
        \cos(l\theta)~~~(l=1,3,5,\ldots),\\
\langle {\bf r}|l,k,-\rangle^{(h)} &\propto& J_l\left({\alpha_{lk}r\over
R}\right)\,
        \sin(l\theta)~~~(l=2,4,6,\ldots),
%\eqno(11)
\end{array}
\end{eqnarray}
where $-\pi/2<\theta<\pi/2$.
Here there is no degeneracy.

For classically nonintegrable cases, 
$\hat{p_\theta}^2$ no longer commutes with $\hat{H_w}$ (but still $[\hat
P,\hat{H_w}]
=0$ for any $w$).
Here Eq.~(8) must be solved numerically.
We use the boundary element method (BEM) [2,18-20] to solve this nonseperable 2D partial 
differential equation. It is an efficient way to solve boundary-value 
problems, because in BEM
a 2D equation with boundary condition becomes an integral equation in
one variable along the boundary.
The method we use is briefly outlined below.

We use a Green's function,
$G({\bf r},{\bf r}')=-{i\over4}\, H^{(1)}_0(k|{\bf r}-{\bf r}'|)$, which
satisfies
\begin{equation}
(\nabla^2+k^2)G({\bf r},{\bf r}')=\delta({\bf r}-{\bf r}'),
%\eqno(12)
\end{equation}
where $H^{(1)}_i(x)$ is the Hankel function of the first kind.
We  multiply Eq.~(8) by $G({\bf r},{\bf r}')$, and multiply Eq.~(12) by
$\Psi({\bf r}')$.
After subtracting one from the other, integrating over the area of $M_w$ 
with respect to ${\bf r}'$, 
and using Green's theorem,
we finally get
\begin{equation}
-\oint_{B_w} ds'\,G({\bf r},{\bf r}')\,u(s')=
        \cases{\Psi({\bf r}), & {\bf r} inside $B_w$,\cr
                {1\over2}\Psi({\bf r}), & {\bf r} on $B_w$,\cr
                0, & {\bf r} outside $B_w$,\cr}
%\eqno(13)
\end{equation}
where $s'$ is arc-length along $B_w=\partial M_w$, $u(s')\equiv{\bf n}'\cdot
\nabla'\Psi({\bf r}(s'))$,
and ${\bf n}'$ is the outward normal unit vector to $B_w$ at ${\bf r}'$.
With ${\bf r}$ lying on $B_w$, by  taking the normal derivative ${\bf n}\cdot
\nabla$ on both sides of Eq.~(13),
we obtain
\begin{equation}
u(s)=-2\oint_{B_w} ds'\,u(s')\,({\bf n}\cdot\nabla) G({\bf r},{\bf r}').
%\eqno(14)
\end{equation}
One way to solve this equation is discretizing $B_w$ by dividing it into $N$
segments. Then Eq.~(14) becomes the matrix equation ${\bf A}\cdot{\bf x}=
{\bf x}$, where ${\bf A}={\bf A}(k)$ is an $(N\times N)$-matrix and ${\bf
x}$ is an
$N$-component vector representing $\{u(s_i)|1\le i\le N\}$.
For given $w$, energy levels of the system, $E_n={\hbar^2}k_n^2/2m$ $(n\ge1)$,  
can be found by solving the equation, $\det|{\bf A}(k)-{\bf I}|=0$.
For each energy level $E_n$, we can obtain $\{u_n(s_i)\}$ by finding
an eigenvector of ${\bf A}(k_n)$ with a near-zero eigenvalue.
Since the numerically obtained energy levels in this way
always have some uncertainty, ``degeneracy" (which is actually
near-degeneracy)
can occur when the difference between two adjacent exact energy levels
is less than the uncertainty.
In these cases, we can find two sets of $\{u(s_i)\}$ with two near-zero 
eigenvalues of ${\bf A}$.
Therefore, looking at eigenvalues of ${\bf A}(k_n)$ can be an easy way to
check numerically for near-degeneracies of an energy level, $E_n$.

For given $w$, we found energy levels $\{ E_n|n\ge1\}$ and normal derivatives,
on the boundary, of
corresponding energy eigenfunctions
$\{\Psi_n({\bf r})\equiv\langle{\bf r}|n\rangle\}$
where $|n\rangle$'s are energy eigenstates.
Then from Eq.~(13) we can calculate the energy eigenfunction inside,
\begin{equation}
\Psi_n({\bf r})={i\over4}\oint_{B_w} ds'\,H^{(1)}_0(k_n|{\bf r}-{\bf r}'|)\,u_n(s').
%\eqno(15)
\end{equation}
Using Eq.~(15),
we can also calculate $\langle {\bf r}|\hat{p_\theta}^2|n\rangle$,
which will be used in calculations of quantum webs in the next subsection,
\begin{equation}
\begin{array}{rcl}
\langle{\bf r}|\hat{p_\theta}^2|n\rangle &=&
                -\hbar^2{\partial^2\over\partial \theta^2}\Psi_n({\bf r})\\
                &=&{i\over4}\oint_{B_w} ds'\,\left[{\partial^2\over
                        \partial \theta^2}H^{(1)}_0(k_n|{\bf r}-{\bf
			r}'|)\right] u_n(s'),
%\eqno(16)
\end{array}
\end{equation}
where
\begin{equation}
\begin{array}{rcl}
{\partial^2\over\partial \theta^2}H^{(1)}_0(k|{\bf r}-{\bf r}')&=&
k \left[{\partial^2\over\partial\theta^2} |{\bf r}-{\bf r}'| \right]
{H^{(1)}_0}'(k|{\bf r}-{\bf r}'|) \\
 &&+ k^2 \left[{\partial\over\partial\theta}|{\bf r}
-{\bf r}'|\right]^2 {H^{(1)}_0}''(k|{\bf r}-{\bf r}'|),
%\eqno(17)
\end{array}
\end{equation}
and we use
\begin{equation}
|{\bf r}-{\bf r}'|=\sqrt{(r\cos\theta-x')^2+(r\sin\theta-y')^2},
%\eqno(18)
\end{equation}
and
\begin{eqnarray}
{H^{(1)}_0}'(x)  &=&-H^{(1)}_1(x),\nonumber\\
          {H^{(1)}_0}''(x) &=&-H^{(1)}_0(x)+{H^{(1)}_1(x)\over x}.
%\eqno(19)
\end{eqnarray}
In calculations of $\Psi_n$ and ${\partial^2\over\partial\theta^2}\Psi_n$,
care must be taken when ${\bf r}$ is close to the boundary because
$H^{(1)}_i(|{\bf r}-{\bf r}'|)$ diverges as $|{\bf r}-{\bf r}'|$
goes to zero.

\subsection*{B. The Quantum Web}
\indent\indent
The quantum-web analysis can be used to observe the manifestations of
classical
chaos in quantum mechanics, and until now has been used primarily in
spin systems [13,14].
Here we will look at three different cases: classically integrable
cases ($w=1,2$), nonintegrable cases showing
soft chaos ($0<w<1$), and nonintegrable cases with hard chaos
($1<w<2$).

For classically integrable  cases ($w=1,2$), we have seen in Sec.~II that
there are
two constants of motion, $E$ and $|p_\theta|$,  and that we can find
two action variables $(J_r,J_\theta)$. 
(There exists a nonlinear map
from $(J_r,J_\theta)$-space to $(E,|p_\theta|)$-space.)
In Fig.~3, a classical orbit appears
as a point in $(E,|p_\theta|)$-space. 
In quantum mechanics, there exist simultaneous
eigenstates of two operators $\hat{H_w}$ ($w=1$ or 2) and $\hat{p_\theta}^2$
[see Eqs.~(10) and (11)]. We can construct a 2D space,
in which a pair of eigenvalues $(E_{lk},|l\hbar|)$ of each eigenstate
is plotted as a point. In Figs.~5(e) and 5(l), we observe the structure of
a deformed lattice. 
This can be understood as a mapping from an almost perfect 2D lattice
in $(J_r,J_\theta)$-space to the deformed lattice in $(E, |p_\theta|)$-space. 
This almost perfect lattice-structure can be explained from
Einstein-Brillouin-Keller (EBK) semiclassical quantization,
\begin{eqnarray}
J_r & \simeq &(n_r-{1\over4})\hbar~~~(n_r:{\rm positive~integer}),\nonumber\\
J_\theta & =&p_\theta=l\hbar~~~(l:{\rm integer}),
%\eqno(20)
\end{eqnarray}
where $J_r$ is a very good approximation and $J_\theta$ is exact.
This is the quantum analogue of Fig.~3. Each point in Fig.~3 represents a
2-torus.
Thus, we can find a 2-torus in classical phase space corresponding
to a quantum eigenstate, and then each eigenstate here corresponds to a
group of
orbits that are on this 2-torus.

For classically nonintegrable cases,
$[\hat{p_\theta}^2,\hat{H_w}]\ne0$
when $w\ne1$ or 2. However, we can still calculate an expectation
value of $\hat{p_\theta}^2$
for an energy eigenstate $|n\rangle$, $\langle n|\hat{p_\theta}^2|n\rangle$,
numerically using Eqs.~(15) and (16),
\begin{eqnarray}
\langle n|\hat{p_\theta}^2|n\rangle
&=&\int_{M_w}d^2{\bf r}\,\langle n|{\bf r}\rangle\langle{\bf r}
        |\hat{p_\theta}^2|n\rangle \nonumber\\
&=&-\hbar^2\int_{M_w}d^2{\bf r}\,\Psi_n^*({\bf r}) {\partial^2\Psi_n({\bf
r})\over
        \partial\theta^2}.
%\eqno(21)
\end{eqnarray}
These values can be interpreted as time-averages of $\hat{p_\theta}^2$ [12].
When there is an accidental degeneracy (or near-degeneracy), we find expectation
values from
eigenvalues of the matrix representation of $\hat{p_\theta}^2$ in the degenerate
subspace.
In this way, we obtain a pair of values $(E_n,\sqrt{\langle n|\hat{p_\theta}^2
|n\rangle})$ for each energy eigenstate. These points
can also be plotted in a 2D space as a quantum web.
We expect that the lattice-structure that exists for integrable cases will be
broken because EBK quantization doesn't apply to nonintegrable
cases.

In Figs.~5(a)-5(d), quantum webs are shown for cases of soft chaos.
When $w=0.99$ [Fig.~5(d)], we observe breaking of the web near conditions
of primary resonances in classical mechanics. We see patterns of crossing
near $(1,4)'$- and $(1,3)'$-resonances. Although ``regular" parts still exist,
some layers seem to start to shift near resonance conditions.
We can roughly estimate the energy value at which the effect starts
to be seen for
each resonance condition by measuring the width of island chains, $\Delta
p_\theta$, in Fig.~4(d).
For example, $(1,2)'$-resonance has the biggest width, and next
$(1,4)'$-resonance, and so on. Because $p_\theta$ is scaled by
$(R\sqrt{2mE})^{-1}$
in Fig.~4, $\Delta p_\theta$ is proportional to $\sqrt{E}$.
When $\Delta p_\theta(=\Delta J_\theta)\geq\hbar$, the resonance
can be clearly seen in the quantum system, and we can roughly obtain an estimate
of the minimum energy at which each resonance is in effect.
The smaller the width of an island chain and the
lower the energy, the less likely to find the web broken near the
curve of the particular resonance. When $w=0.7$ [Fig.~5(c)], we see similar
patterns as in Fig.~5(d).
When $w=0.5$ [Fig.~5(b)], the classical system has a large chaotic region 
in the PSS, and has 
periodic orbits with neutral stability, which reside on two axes, 
$\theta=0$ and $p_\theta=0$.
The quantum web, however, is
quite regular although the structure looks different from those of integrable cases.
It looks more like a structure of layers.
When $w=0.3$ [Fig.~5(a)], the web is similar to that of
Fig.~5(b).

In Figs.~5(f)-(k), we show quantum webs for cases of hard chaos.
When $w=1.01$ [Fig.~5(f)], the lattice-structure is still intact except
a little kink, although this is the fully chaotic case classically [see
Fig.~4(a)].
We observe the lattice-structure quickly collapses as we increase $w$.
When $w=1.5$ [Fig.~5(i)], the structure is very irregular except for four
regularly placed points 
near the top-right corner. 
(Some of eigenstates noted by arrows here will be
examined in the next subsection using Husimi plots.)
The case of $w=1.7$ [Fig.~5(j)] is the most irregular quantum web among
cases shown.
When $w=1.9$ [Fig.~5(k)], we observe splitting of degeneracies and also
quick collapse
of lattice-structure from an integrable case $w=2$ [Fig.~5(l)].
As we have seen so far, the lattice structure tends to collapse quickly
irrespective of energy in cases of hard chaos, but there also exist
small remnants of regularity in some cases.

\subsection*{C. Quantum Poincar\'{e} Section}
\indent\indent
The Husimi plot provides a method of extracting
a quantum Poincar\'{e} surface of section (QPS) from a quantum state [15].
The QPS is a quantum analogue of PSS, which we have obtained in Sec.~II.
The Husimi function of an 1D system corresponding to a state $|\Psi\rangle$
is defined as
\begin{equation}
F(x_0,p_0)=\left| \langle x_0,p_0 | \Psi\rangle \right|^2,
%\eqno(22)
\end{equation}
where $|x_0,p_0\rangle$ is a coherent state with a representation in
configuration space,
\begin{equation}
\langle x|x_0,p_0\rangle=\left({1\over\pi\sigma^2}\right)^{1\over4}\,
\exp\left[-{(x-x_0)^2\over2\sigma^2}+{i\over\hbar}p_0(x-x_0)\right].
%\eqno(23)
\end{equation}
In 2D billiards, Eq.~(22) can be modified to create a Husimi
function using the coordinate along the boundary [16].
For example, for $M_w$-billiard along the circular part of the boundary,
the Husimi function is defined as
\begin{equation}
F_n(\theta_0,p_{\theta 0})=\left|
\int^{\theta_{max}}_{-\theta_{max}} d\theta'\,
\langle\theta_0,p_{\theta 0}|\theta'\rangle\, u_n(\theta') \right|^2,
%\eqno(24)
\end{equation}
where $-\theta_{max}<\theta_0<\theta_{max}$ and
$-R\sqrt{2mE}<p_{\theta0}<R\sqrt{2mE}$.
Here $\langle\theta'|\theta_0,p_{\theta0}\rangle$ has the same form as in Eq.~(23)
with $\sigma$ given by the value $[{\theta_{max}\hbar/(R\sqrt{2mE})}]^{1/2}$,
and $u_n(\theta)$ is the normal derivative of the energy eigenfunction
on the circular part of the boundary,
${\partial\over\partial r}\Psi_n({\bf r})|_{r=R}$.

In Fig.~6, we show Husimi plots for selected eigenstates for three cases
($w=0.5,0.9,1.5$). And, in Fig.~7, we show probability densities of wave
functions, $|\Psi_n({\bf r})|^2$, for some of eigenstates chosen from Fig.~6.
(Most eigenstates chosen in Fig.~6 can be found in Figs.~5(b),5(d), and 5(i)
with arrows pointed to them.)

Figures 6(a)-6(d) show Husimi plots of energy eigenstates when $w=0.5$.
The pattern of PSS shown in Fig.~4(f) can be seen in these plots.
We can get some information on chosen eigenstates from the quantum web [Fig.~5(b)].
The eigenstate for Fig.~6(a) is on the outer part, and the eigenstate for
Fig.~6(b) is on the inner part of the quantum web.
Figure 6(a) shows chaotic region of PSS, and Fig.~6(b) seems to correspond
with a two-bounce orbit with neutral stability, which we can observe in
the wave function [Fig.~7(a)]. The eigenstate for Fig.~6(c) is in the middle of
the quantum web, and its Husimi plot and wave function [Fig.~7(b)] lie
between two extreme cases above. The eigenstate for Fig.~6(d) has
a relatively high energy, but the structure is similar to
Fig.~6(c).

Figures 6(e)-6(h) show Husimi plots of energy eigenstates when $w=0.9$,
and each eigenstate is picking up a classical primary resonance shown
as an island chain in PSS [Fig.~4(d)].
The eigenstate for Fig.~6(e), located near the $(1,2)'$-resonance in the quantum web
[Fig.~5(d)], shows the pattern of the island chain of the $(1,2)'$-resonance,
although it is only the 9th highest energy eigenstate.
The eigenstates of Figs.~6(f) and 6(g), located at the crossing of two
layers near the
$(1,4)'$-resonance,  show the pattern of the island chain
of the $(1,4)'$-resonance.
Wave functions [Figs.~7(c) and 7(d)] of these states show the trace of
unstable and stable periodic orbits [see Fig.~2(b)], respectively.
The eigenstate for Fig.~6(h), which is on the $(1,3)'$-resonance,
also shows the pattern of the island chain of the $(1,3)'$-resonance.
As expected earlier, the $(1,2)'$-resonance is observed in the Husimi plot
at lower-energy eigenstates than the $(1,4)'$-resonance.

Figures 6(i)-6(l) show Husimi plots of energy eigenstates when $w=1.5$.
The eigenstate for Fig.~6(i), located in the inner part of the quantum web
[Fig.~5(i)], seems to be picking up the periodic orbit with neutral stability,
which can be seen clearly in the wave function [Fig.~7(e)].
The eigenstate of Fig.~6(k), which is one of four regularly placed points in the
quantum web, shows a {\em whispering} {\em gallery} state [Fig.~7(f)].
All four of these regularly placed eigenstates show similar Husimi plots.
A relatively high energy eigenstate for Fig.~6(l)
shows a more uniformly distributed Husimi plot like the corresponding 
PSS in classical mechanics.

%-----------------------------------------------------------------
\section*{IV. Conclusions}
\indent\indent
The quantum web is the quantum representation of
$(E,|p_\theta|)$-space in Fig.~3. In the regime of soft chaos,
we observe that the lattice-structure obtained for the integrable case starts
to break near the primary resonance conditions obtained
from classical mechanics as the width parameter, $w$, decreases from $w=1$.
The effect of resonances is greater
when energy is higher and the width of an island chain in the PSS is  greater.
Even for the most chaotic case in the soft-chaos regime, layer-structure
remains.
In the hard-chaos regime, the regular quantum web
collapses more quickly regardless of energy as the width parameter, $w$,
increases from $w=1$.

%-----------------------------------------------------------------
\section*{Acknowledgments}
\indent\indent
The authors wish to thank the Welch Foundation, Grant
No.F-1051, NSF Grant No.INT-9602971, and DOE contract No.DE-FG03-94ER14405
for partial support of this work.
We thank NPACI and the University of Texas High Performance Computing Center for use
of its
facilities.

%-----------------References------------------------------

\end{multicols}

\twocolumn
%--------list of figures----------
\begin{figure}[h]
 \begin{center}
	 \scalebox{0.45}[0.45]{\includegraphics{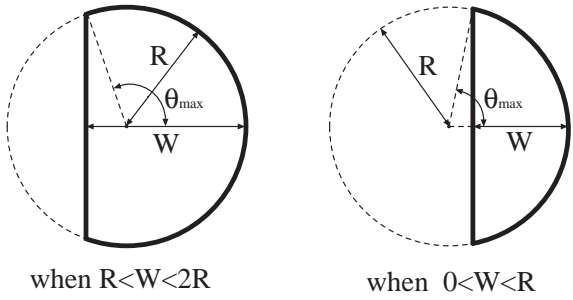}}
 \end{center}
	\caption{
		Geometry of the ``Moon" billiard. When $W=2R$, it is a full
		circle, and when $W=R$, it is a half circle. Here $\theta_{max}$
		is given by the equation, $\cos \theta_{max}=(R-W)/R$.
		}
\end{figure}

\begin{figure}[h]
 \begin{center}
	 \scalebox{0.45}[0.45]{\includegraphics{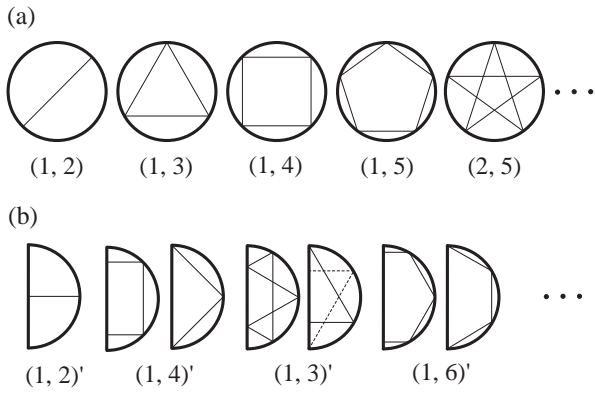}}
 \end{center}
	\caption{
		Closed orbits in integrable cases. 
		(a) In a full circle, periodic orbits can be classified as 
		$(m,n)$ where orbits have $m$ bounces and $n$ revolutions
		in a cycle. (b) In a half circle, we can use the notation
		of a full circle to classify periodic orbits as $(m,n)'$.
		}
\end{figure}
\begin{figure}[h]
 \begin{center}
	 \scalebox{0.45}[0.45]{\includegraphics{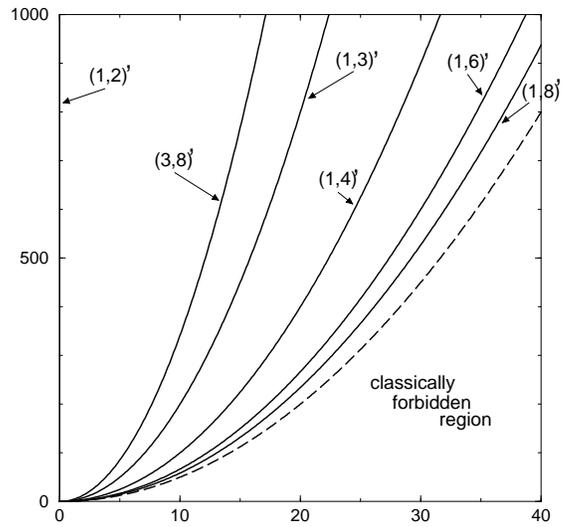}}
 \end{center}
	\caption{
		For a half circle, the condition in $(E,|p_\theta|)$-space
		to have periodic orbits $(m,n)'$ is given by Eq.~(5), and some 
		are plotted. 
		These curves
		are densely populated in classically allowed region.
		}
\end{figure}

\clearpage
\begin{figure}[h]
	\caption{
		The Poincar\'{e} surface of section of $M_w$-billiard varying $w$
		where $w=W/R$. We observe ergodic motions when $1<w<2$, and
		generic chaotic behaviors when $0<w<1$.
		(a) Ergodic when $w=1.01$ with one orbit.
		(b) Integrable when $w=1$.
		(c) $w=0.99$. (d) $w=0.9$. (e) $w=0.7$.
		(f) $w=0.5$. Closed orbits residing in
		$(\theta=0)$-axis and $(p_\theta=0)$-axis have neutral stabilities.
		One orbit is filling almost all space except two axes.
		(g) $w=0.3$. (h) $w=0.1$.
		}
\end{figure}
\begin{figure}
	\caption{
		The quantum web of $M_w$-billiard varying $w$. Even and odd parity eigenstates
		are shown together.
		(a) $w=0.3$.
		(b) $w=0.5$.
		(c) $w=0.7$.
		(d) $w=0.9$. The condition of primary resonances in classical mechanics
			are also shown.
		(e) $w=1.0$. Classically integrable case.
		(f) $w=1.01$.
		(g) $w=1.1$.
		(h) $w=1.3$.
		(i) $w=1.5$.
		(j) $w=1.7$.
		(k) $w=1.9$.
		(l) $w=2.0$. Classically integrable case. There are two-fold
			degeneracies when $p_\theta\ne0$.
		[Arrows in (b), (d), and (i) indicate states which will be studied
			using Husimi plots in Fig.~6.]
		}
\end{figure}
\begin{figure}
	\caption{
		Husimi plots for given eigenstates. 
		($\epsilon\equiv E/({\hbar^2\over mR^2})$)
		(a) $w=0.5$ and $\epsilon=143.28$.
		(b) $w=0.5$ and $\epsilon=740.79$.
		(c) $w=0.5$ and $\epsilon=1156.07$.
		(d) $w=0.5$ and $\epsilon=9992.23$.
		(e) $w=0.9$ and $\epsilon=62.513$.
		(f) $w=0.9$ and $\epsilon=365.64$.
		(g) $w=0.9$ and $\epsilon=371.99$.
		(h) $w=0.9$ and $\epsilon=617.98$.
		(i) $w=1.5$ and $\epsilon=258.03$.
		(j) $w=1.5$ and $\epsilon=284.94$.
		(k) $w=1.5$ and $\epsilon=268.80$.
		(l) $w=1.5$ and $\epsilon=842.69$.
		[A square at the top right corner of each plot represents
		the size of $h$ (Planck constant). 
		All eigenstates except (d) and (l) are pointed in quantum webs in
			Fig.~5 by arrows.]
		}
\end{figure}
\begin{figure}
	\caption{
		The probability density of some 
		energy eigenstates.
		($\epsilon\equiv E/({\hbar^2\over mR^2})$)
		(a) $w=0.5$ and $\epsilon=740.79$. [See Fig.~6(b).]
		(b) $w=0.5$ and $\epsilon=1156.07$. [See Fig.~6(c).]
		(c) $w=0.9$ and $\epsilon=365.64$. [See Fig.~6(f).]
		(d) $w=0.9$ and $\epsilon=371.99$. [See Fig.~6(g).]
		(e) $w=1.5$ and $\epsilon=258.03$. [See Fig.~6(i).]
		(f) $w=1.5$ and $\epsilon=268.80$. [See Fig.~6(k).]
		}
\end{figure}

\end{document}